\documentclass[11pt,english,superscriptaddress,showpacs,aps,preprint]{revtex4}
\usepackage[latin9]{inputenc}
\usepackage{amsmath}
\usepackage{amssymb}
\usepackage{graphicx}
\usepackage{esint}

\makeatletter
\@ifundefined{textcolor}{}
{%
 \definecolor{BLACK}{gray}{0}
 \definecolor{WHITE}{gray}{1}
 \definecolor{RED}{rgb}{1,0,0}
 \definecolor{GREEN}{rgb}{0,1,0}
 \definecolor{BLUE}{rgb}{0,0,1}
 \definecolor{CYAN}{cmyk}{1,0,0,0}
 \definecolor{MAGENTA}{cmyk}{0,1,0,0}
 \definecolor{YELLOW}{cmyk}{0,0,1,0}
}

\usepackage{slashed}

\def\eff{\mathrm{eff}}

\usepackage{babel}

\makeatother

\usepackage{babel}
\begin{document}

\title{Generation of Axion-Like Couplings via Quantum Corrections in a Lorentz
Violating Background}

\author{L. H. C. Borges}

\email{luizhenriqueunifei@yahoo.com.br}

\author{A. G. Dias}

\email{alex.dias@ufabc.edu.br}

\author{A. F. Ferrari}

\email{alysson.ferrari@ufabc.edu.br}

\affiliation{Universidade Federal do ABC, Centro de Ciências Naturais e Humanas,
Rua Santa Adélia, 166, 09210-170, Santo André, SP, Brasil}

\author{J. R. Nascimento}

\email{jroberto@fisica.ufpb.br}

\author{A. Yu. Petrov}

\email{petrov@fisica.ufpb.br}

\affiliation{Departamento de Física, Universidade Federal da Paraíba\\
 Caixa Postal 5008, 58051-970, João Pessoa, Paraíba, Brazil}
\begin{abstract}
Light pseudoscalars, or axion like particles (ALPs), are much studied
due to their potential relevance to the fields of particle physics,
astrophysics and cosmology. The most relevant coupling of ALPs from
the viewpoint of current experimental searches is to the photon: in
this work, we study the generation of this coupling as an effect of
quantum corrections, originated from an underlying Lorentz violating
background. Most interestingly, we show that the interaction so generated
turns out to be Lorentz invariant, thus mimicking the standard ALPs
coupling to the photon that is considered in the experiments. This
consideration implies that violations of spacetime symmetries, much
studied as possible consequences of physics in very high energy scales,
might infiltrate in other realms of physics in unsuspected ways. Additionally,
we conjecture that a similar mechanism can also generate Lorentz invariant
couplings involving scalar particles and photons, playing a possible
role in the phenomenology of Higgs bosons. 
\end{abstract}


\maketitle

\section{\label{sec:Introduction}Introduction}

Theoretical motivations for the introduction of light pseudoscalars
in the Standard Model ranges from solving specific technical issues
in the theory of strong interactions to more general questions such
as the search for dark matter candidates (for a theoretical introduction
and some experimental developments, see for example\,\cite{Jaeckel:2010ni,Ringwald:2012hr}).
The search for these light pseudoscalars has been an active field
of experimental work, usually by looking for the interaction between
these particles and the photon. Our main purpose in this work is to
show that this particular interaction can be generated from an underlying
Lorentz violating background. More precisely, we will assume the existence
of a massive fermionic field with Lorentz violating (LV) interactions
of a specific form, and show that the resulting low energy effective
Lagrangian contains Lorentz invariant (LI) interactions whose intensity
are functions of LV parameters. We will argue that these LI terms
are the dominant ones in the low energy phenomenology of pseudoscalars
interacting with photons, thus proposing a scenario where the most
relevant effects of the LV are actually ``standard'' LI interactions.
In this way the experiments for searching light pseudoscalars would
also be able to probe the specific setup of LV that we consider --
conversely, our result suggests that, contrary to common belief, an
underlying LV could be relevant to the study of typically LI phenomena.
This finding is relevant in pointing out that possible violations
of spacetime symmetries could permeate other areas of physics, with
possible consequences in contexts where such violations are not typically
considered relevant. 

The axion is a well studied example of light pseudoscalar appearing
as a pseudo-Nambu-Goldstone boson from the breaking of an anomalous
global chiral $U\left(1\right)$ symmetry, the Peccei-Quinn (PQ) symmetry,
and leading to a dynamical solution for the strong CP problem in QCD\,\cite{Peccei:1977hh,Weinberg:1977ma,Wilczek:1977pj}.
Experiments for searching the axion~\cite{Sikivie:1983ip}, both
direct and via astrophysical observations, are being performed since
the 1980's, and a class of \textquotedblleft invisible axion\textquotedblright -type
models\,\cite{Kim:1979if,Shifman:1979if,1981PhLB..104..199D,Zhitnitsky:1980tq}
are still phenomenologically viable. More concretely, consider a pseudoscalar
field $\phi$ coupled to a charged fermion field $\psi$ according
to the term $i\phi\overline{\psi}\gamma_{5}\psi$. Assuming that the
fermionic field is heavy it can be integrated out, and the low energy
effective Lagrangian describing the pseudoscalar particles interacting
with photons turns out to be 
\begin{equation}
\mathcal{L}_{{\rm eff}}=-\frac{1}{4}F_{\mu\nu}F^{\mu\nu}+\frac{1}{2}\partial_{\mu}\phi\partial^{\mu}\phi-\frac{m_{\phi}^{2}}{2}\phi^{2}-\frac{g_{\phi\gamma}}{4}\phi F_{\mu\nu}\widetilde{F}^{\mu\nu}\,,\label{eq:leff}
\end{equation}
where $g_{\phi\gamma}$ is a coupling constant with inverse of mass
dimension, $m_{\phi}$ is the axion mass, $F_{\mu\nu}$ is the electromagnetic
field-strength, and $\tilde{F}^{\mu\nu}=\varepsilon^{\mu\nu\alpha\beta}F_{\alpha\beta}$
its dual. In axion models there is a specific relation between $m_{\phi}$
and $g_{\phi\gamma}$, both related to the axion decay constant $f_{a}$
which is assumedly much higher than the electroweak scale. Indeed\,\cite{Beringer:1900zz},
\begin{equation}
m_{\phi}=\frac{z^{1/2}}{1+z}\frac{f_{\pi}m_{\pi}}{f_{a}}\,;\, g_{\phi\gamma}=\frac{\alpha}{2\pi f_{a}}\left(r-\frac{2}{3}\frac{4+z}{1+z}\right)\,,\label{eq:mAgArelation}
\end{equation}
where $m_{\pi}$ and $f_{\pi}$ are the pion mass and decay constant,
$z$ is the ratio of the masses of up and down quarks, $\alpha$ is
the QCD coupling constant, and $r$ is the ratio of the electromagnetic
and color anomalies of the axial current associated with PQ symmetry.
Despite the ratio $r$ being model dependent to some degree, this
relation is a typical signature of the light pseudoscalar related
to the PQ mechanism, which is conventionally denoted the axion. 

More generically, particle excitations of very light pseudoscalar
fields having interactions like the one in Eq.\,\eqref{eq:leff}
are denoted as axion-like-particles (ALPs), and they are not necessarily
connected to the properties of the QCD vacuum. For example, compactifications
in string theory have been shown to produce light pseudoscalars\,\cite{Svrcek:2006yi}
(see also\,\cite{Ringwald:2012cu} and references therein). 

The last term in Eq.~\eqref{eq:leff} is the lowest dimensional operator
for the pseudoscalar field interaction with the electromagnetic field.
Therefore, the interaction 
\begin{equation}
\mathcal{L}_{{\rm int}}=-\frac{g_{\phi\gamma}}{4}\phi F_{\mu\nu}\tilde{F}^{\mu\nu}=g_{\phi\gamma}\phi\vec{E}\cdot\vec{B}\,,\label{eq:axion-photon-coupling}
\end{equation}
is the most relevant one in the low energy regime. It leads to the
remarkable phenomenon of ALP-photon oscillation: even with feeble
couplings with the usual matter, an hypothetical ALP of very low mass
can oscillate into a detectable photon when passing through a magnetic
field\,\cite{Sikivie:1983ip}. Combining the up to date experimental
results and the constraints from energy loss in stars\,\cite{raffeltbook},
the interaction in Eq.~\eqref{eq:axion-photon-coupling} is now excluded
in an impressive range of values $g_{\phi\gamma}\gtrsim10^{-10}\mbox{GeV}^{-1}$
for a wide range of mass $m_{\phi}$ in the sub-eV scale\,\cite{Beringer:1900zz}.
Also, new proposals for searching ALPs whose couplings with photons
could be directly tested up to $g_{\phi\gamma}\sim10^{-11}-10^{-12}\mbox{GeV}^{-1}$
are underway\,\cite{Vogel:2013bta,Bahre:2013ywa}. 

For all of this the interaction in Eq. \eqref{eq:axion-photon-coupling}
is presently one of the most important proposals of new physics relevant
to the low energy regime, originated from some theory associated to
a high mass scale parameter $M$, such that $g_{\phi\gamma}\propto1/M$.
The models from which Eq.\,\eqref{eq:axion-photon-coupling} is derived
are generically Lorentz invariant (LI). In this work, we follow this
perspective, but relaxing the condition of Lorentz invariance: we
consider some Lorentz violating new physics at a very high energy
scale, and look for effects which are relevant to a very low energy
phenomenology.,We will show a specific setup of LV that can generate
the interaction in Eq.\,\eqref{eq:axion-photon-coupling} as its
dominant effect. The LV is assumed here to occur due to the existence
of constant vectors $b^{\mu},\, d^{\mu}$ in the interaction of the
pseudoscalar and electromagnetic fields with the heavy fermion, which
after being integrated out will lead to $g_{\phi\gamma}\propto b\cdot d$. 

Concerning theories which include LV, a framework for the incorporation
of a possible LV within our current understanding of the elementary
interactions, called Standard Model Extension (SME), has been used
to provide an environment for testing the validity of relativistic
symmetry in different phenomena\,\cite{colladay:1998fq,Kostelecky:2008ts,mattingly:2005re}.
In essence, one incorporates in the Standard Model Lagrangian all
possible covariant, renormalizable and gauge invariant terms that
involve constant background tensors. These are assumed to appear from
a more fundamental theory at very high energy scales -- via a spontaneous
symmetry breaking that gives a non vanishing vev for some tensorial
field, for instance\,\cite{kostelecky1989spontaneous}. One example
is the so-called Carrol-Field-Jackiw (CFJ) term $k_{\mu}\epsilon^{\mu\nu\rho\sigma}A_{\nu}\partial_{\rho}A_{\sigma}$\,\cite{carroll:1989vb},
which leads to modified propagation of light waves in vacuum and --
since these effects are not seen in our current experiments -- to
strong bounds on the value of the Lorentz violating vector $k_{\mu}$.
The CFJ term can be induced by quantum corrections, starting from
the Lorentz breaking coupling $b_{\mu}\overline{\psi}\gamma_{5}\gamma^{\mu}\psi$
and integrating out the fermionic degrees of freedom, thus providing
a connection between the $b_{\mu}$ and the vector $k_{\mu}$; it
allows the translation of constraints from the photon sector to other
sectors of the SME, strongly constraining these models. This mechanism
have been intensively studied in recent years (see\,\cite{Brito:2007uc}
and references therein, and also\,\cite{Scarp2013,Mariz:2011ed,Gomes:2009ch}),
in part due to the appearance of regularization dependent integrals
in the calculations, an aspect that will also appear in our model.
However, it must be pointed out that while the CFJ term is quadratic,
directly modifying the photon propagation in vacuum, in our context
the LV affects only the interactions of the photon with the ALP field.

The SME have been extended to include gravitational interactions\,\cite{kostelecky-gravityLV}
and, more recently, its extension to higher dimensional operators
is being worked out\,\cite{kostelecky-QED-higherdimension-2009,kostelecky-fermionshigherdimension-2013}.
Our model is also in some sense an extension of the SME, since its
starting point is a model containing LV interactions involving Standard
Model fields and two new ingredients: the light pseudoscalar, whose
low energy phenomenology we shall be interested in unveiling, and
a massive fermion whose integration in the effective action will provide
the translation of the very high energy LV to a low energy LI effect. 

This work is organized as follows. The model we shall be dealing with
is presented in Section \ref{sec:The-Model}, involving a new massive
fermion and a pseudoscalar fields coupled via LV interactions to the
gauge fields of the Standard Model. In Section \ref{sec:Generation-of-ALPs-Photon},
we calculate the corrections to the effective action generated after
integration of the heavy fermion, and show that the dominant operator
generated by the LV in the low energy regime is LI, and reproduces
the standard ALP-photon interaction. We comment on possible consequences
of this result for ALP phenomenology in Section \ref{sec:Implication-for-ALPs}.
Finally, Section \ref{sec:Conclusions-and-Perspectives} contains
our conclusions and some final remarks.

\section{\label{sec:The-Model}The Model}

We now show explicitly how the low energy effective interaction between
the pseudoscalar and the electromagnetic fields can be generated from
a model for high energy physics which has LV interactions. Let us
take into account a single charged vectorial fermionic field $\psi\sim\left(\mathbf{1},\,1\right)$,
and the pseudoscalar field $\phi\sim\left(\mathbf{1},\,0\right)$,
where the numbers in parenthesis means the transformation under the
Standard Model gauge group $SU\left(2\right)_{L}\otimes U\left(1\right)_{Y}$.
Besides the minimal coupling with the hypercharge $U\left(1\right)_{Y}$
gauge field $B_{\mu}$, it is assumed that $\psi$ has a LV interaction
$B_{\mu\nu}d^{\nu}\overline{\psi}\gamma^{\mu}\psi$, where $B_{\mu\nu}=\partial_{\mu}B_{\nu}-\partial_{\nu}B_{\mu}$.
Also, the pseudoscalar field $\phi$ enters via a LV Yukawa coupling
$\bar{\psi}\gamma_{5}\slashed{b}\phi\psi$. The starting Lagrangian
is therefore 
\begin{equation}
{\cal L}=\bar{\psi}\left[i\slashed{\partial}-m-\gamma^{\mu}(g^{\prime}B_{\mu}+B_{\mu\nu}d^{\nu})-\gamma_{5}\slashed{b}\phi\right]\psi\,,\label{eq:Lagrangian}
\end{equation}
where the fermion mass $m$ is supposedly higher than the electroweak
scale, actually high enough to justify the integration of its degrees
of freedom. Also, $g^{\prime}$ is the gauge coupling constant, and
$b^{\mu}$, $d^{\mu}$ are constant vectors leading to LV; it is worth
mentioning that $b^{\mu}$, $d^{\mu}$ have different mass dimensions.
The model \eqref{eq:Lagrangian} is an extension of the one used earlier
in \cite{2012JPhG...39c5002G,Mariz:2011ed,Scarp2013} for studies
of the perturbative generation of the aether term; in\,\cite{Gomes:2009ch}
its version involving only the non minimal interaction was used for
the same purpose. 

We shall not be interested in investigating the origin of the LV terms,
but rather to show that from Eq.\,\eqref{eq:Lagrangian} we can generate
the interaction in Eq.\,\eqref{eq:axion-photon-coupling}, after
the field $\psi$ is integrated out. Notice that we do not have to
consider the $\phi$ kinetic term in our calculations, that means
the mechanism we shall describe is completely independent of the mass
of pseudoscalar, which can be taken to be as small as necessary to
fit experimental constraints. 

The one-loop correction to the effective action of the gauge field
$B_{\mu}$ can be expressed as usual in terms of a functional trace,
\begin{equation}
S_{\eff}\left[B\right]=-i\,{\rm Tr}\,\ln\left(i\slashed\partial-m-\gamma^{\mu}\tilde{B}_{\mu}-\gamma_{5}\slashed{b}\phi\right)\,,\label{eq:Seff1}
\end{equation}
where 
\begin{equation}
\tilde{B}_{\mu}=g^{\prime}B_{\mu}+B_{\mu\nu}d^{\nu}.\label{eq:Atilde}
\end{equation}
This effective action can be expanded in the following power series,
\begin{equation}
S_{\eff}\left[B\right]=i\,{\rm Tr}\sum_{n=1}^{\infty}\frac{1}{n}\Biggl[\frac{1}{i\slashed\partial-m}\,\left(\gamma^{\mu}\tilde{B}_{\mu}+\gamma_{5}\slashed{b}\phi\right)\Biggr]^{n}.\label{eq:Seff2}
\end{equation}

From Eq.\,\eqref{eq:Seff2} we will obtain a myriad of effective
interactions involving the pseudoscalar and gauge fields. On general
grounds, only the lowest dimensional ones are expected to contribute
to the low energy phenomenology we are interested in, since the others
will involve additional powers of the large fermion mass $m$ in the
denominator. We will argue, in the following Section, that the dominant
interactions generated in the ALP-photon sector are LI, and reproduce
in form Eq.\,\eqref{eq:axion-photon-coupling}.

\section{\label{sec:Generation-of-ALPs-Photon}Generation of ALPs-Photon Interaction}

\begin{figure}
\begin{centering}
\includegraphics[scale=0.4]{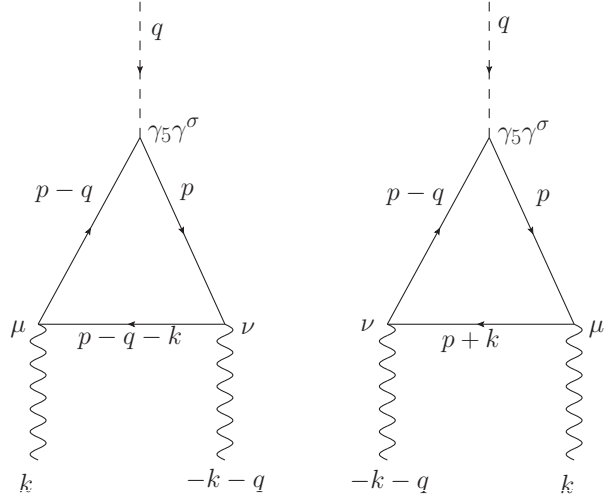} 
\par\end{centering}

\protect\caption{\label{fig:fig1}Triangle graphs corresponding to Eq.\,\eqref{eq:PolarizationTensor1};
the momenta associated with the external lines are momenta \emph{entering
}the corresponding line.}
\end{figure}

First we show how to extract the ALP-photon interaction vertex from
Eq\,\eqref{eq:Seff2}, after that we will justify the fact that this
interaction is indeed the dominant one in the low energy regime. We
start by isolating in Eq.~\eqref{eq:Seff2} the contributions of
the second order in $\tilde{B}_{\mu}$ and first order in $\phi$,
\begin{equation}
S_{\eff}\left[B\right]\supset i\,{\rm Tr}\Biggl[\frac{1}{i\slashed\partial-m}\gamma^{\mu}\tilde{B}_{\mu}\frac{1}{i\slashed\partial-m}\gamma^{\nu}\tilde{B}_{\nu}\frac{1}{i\slashed\partial-m}\gamma_{5}\slashed{b}\phi\Biggr]\,,\label{eq:Seff2b}
\end{equation}
where the cyclic property of the trace have been used. Taking into
account Eq.\,\eqref{eq:Atilde}, after Fourier transform, one obtains
\begin{equation}
S_{\eff}\left[B\right]\supset g^{\prime}d^{\rho}b_{\sigma}\int\frac{d^{4}k}{\left(2\pi\right)^{4}}\frac{d^{4}q}{\left(2\pi\right)^{4}}\,\tilde{\phi}\left(q\right)B_{\mu}\left(k\right)B_{\nu\rho}\left(-k-q\right)\Pi^{\mu\nu\sigma}\left(k,q\right)\,,\label{eq:Seff3}
\end{equation}
where the one-loop correction to the vacuum polarization reads 
\begin{align}
\Pi^{\mu\nu\sigma}\left(k,q\right)\,=\, & i\,{\rm tr}\int\frac{d^{4}p}{(2\pi)^{4}}\gamma^{\mu}S\left(p-k-q\right)\gamma^{\nu}S\left(p\right)\gamma_{5}\gamma^{\sigma}S\left(p-q\right)\nonumber \\
 & +i\,{\rm tr}\int\frac{d^{4}p}{(2\pi)^{4}}\gamma^{\nu}S\left(p+k\right)\gamma^{\mu}S\left(p\right)\gamma_{5}\gamma^{\sigma}S\left(p-q\right)\,.\label{eq:PolarizationTensor1}
\end{align}
Here, $S(p)=(\slashed{p}-m)^{-1}$ is the free Fermion propagator.
Eq. \eqref{eq:PolarizationTensor1} corresponds to the sum of triangle
graphs depicted in Fig.\,\ref{fig:fig1}: it will prove convenient
to calculate it by expanding around $q=0$. Noticing that $\Pi^{\mu\nu\sigma}\left(k,q\right)$
is defined by a potentially linearly divergent integral, we can write
that 
\begin{equation}
\Pi^{\mu\nu\sigma}\left(k,q\right)=\Pi^{\mu\nu\sigma}\left(k,0\right)+q_{\sigma}\frac{\partial}{\partial q_{\sigma}}\left.\Pi^{\mu\nu\sigma}\left(k,q\right)\right|_{q=0}+\mbox{finite terms}\,.\label{eq:ourPi-1}
\end{equation}

The integrals contributing to $\Pi^{\mu\nu\sigma}\left(k,0\right)$
turn out to be rather similar to the ones considered in the perturbative
generation of the CFJ term as studied, for example, in\,\cite{Jackiw:1999yp},
with a difference of momentum routing that amounts to a shift in the
integration momentum $p\rightarrow p-k$. Since these integrals are
linearly divergent, such a shift would produce a \emph{finite} shift
in the result of the integrals, i.e., 
\begin{equation}
\Pi^{\mu\nu\sigma}\left(k,0\right)=\tilde{\Pi}^{\mu\nu\sigma}\left(k\right)+C^{\prime\prime\prime}\varepsilon^{\mu\nu\sigma\rho}k_{\rho}\,,\label{eq:relOursCFJ}
\end{equation}
where $C^{\prime\prime\prime}$ is a finite constant, and $\tilde{\Pi}^{\mu\nu\sigma}\left(k\right)$
is the polarization tensor appearing in\,\cite{Jackiw:1999yp}. The
Feynman integral involved in $\tilde{\Pi}^{\mu\nu\sigma}\left(k\right)$
turns out to be finite and ambiguous upon regularization, but also
proportional to $\varepsilon^{\mu\nu\sigma\rho}k_{\rho}$. We can
therefore conclude that 
\begin{equation}
\Pi^{\mu\nu\sigma}\left(k,0\right)=C^{\prime\prime}\varepsilon^{\mu\nu\sigma\rho}k_{\rho}\,,\label{eq:PolarizationTensor2}
\end{equation}
where the exact value of the \emph{finite }coefficient $C^{\prime\prime}$
depends on the regularization scheme and the way the integral is manipulated.
This ambiguity in the evaluation of triangle graphs is well known,
playing a key role in the study of anomalies in chiral gauge theories,
for example. In that context, ambiguous integrals are fixed by imposing
a physical requirement, such as the conservation of the electric current.
Different physical requirements have also been considered in the calculation
of the CFJ term (see for example\,\cite{chen1999,Andrianov2002,piguet2010}),
but none of these results apply directly to our case. We remind that
the CFJ term is a quadratic term in the effective photon Lagrangian,
while we are studying the generation of an interaction term involving
photons and pseudoscalars. 

In summary, from Eqs.\,\eqref{eq:PolarizationTensor1} and\,\eqref{eq:PolarizationTensor2},
we can state that 
\begin{equation}
\Pi^{\mu\nu\sigma}\left(k,q\right)=C^{\prime\prime}\epsilon^{\mu\nu\sigma\rho}k_{\rho}+q_{\sigma}\frac{\partial}{\partial q_{\sigma}}\left.\Pi^{\mu\nu\sigma}\left(k,q\right)\right|_{q=0}+\mbox{finite terms}\,.\label{eq:ourPi}
\end{equation}
The first term in Eq.\,\eqref{eq:ourPi} is finite and ambiguous;
the second one could be at the most logarithmically divergent by power
counting, but after explicit calculation it turns out to be finite.
In fact, it can be shown that
\begin{align}
q_{\sigma}\frac{\partial}{\partial q_{\sigma}} & \left.\Pi^{\mu\nu\sigma}\left(k,q\right)\right|_{q=0}=\frac{C^{\prime\prime}}{2}\varepsilon^{\mu\nu\sigma\rho}q_{\rho}\nonumber \\
 & +D\left(k^{\mu}\varepsilon^{\nu\sigma\rho\theta}q_{\rho}k_{\theta}+\left(\mu\leftrightarrow\nu\right)\right)+E\varepsilon^{\mu\nu\sigma\rho}k_{\rho}\left(k\cdot q\right)\,,\label{eq:ourPi2}
\end{align}
where again one obtains a contribution from the same finite (yet ambiguous)
Feynman integrand that contributed to the first term in Eq.\,\eqref{eq:ourPi}.
Thus the quantum corrections to the photon effective action in our
theory are completely finite. 

Focusing at the moment on the terms proportional to $\varepsilon^{\mu\nu\sigma\rho}q_{\rho}$
in Eqs.\,\eqref{eq:ourPi} and \eqref{eq:ourPi2}, one can go back
to Eq.\,\eqref{eq:Seff3}, thus obtaining 
\begin{equation}
S_{\eff}\left[B\right]\supset C^{\prime}\, g^{\prime}\,\phi\epsilon^{\rho\mu\nu\lambda}b_{\lambda}d^{\kappa}B_{\rho\mu}B_{\kappa\nu}\,.\label{eq:Seff4}
\end{equation}

After electroweak symmetry breaking, the $B_{\mu}$ field rotates
into a superposition of the photon $A_{\mu}$ and the neutral weak
gauge boson $Z_{\mu}$, 
\begin{equation}
B_{\mu}=\sin\theta\, Z_{\mu}+\cos\theta\, A_{\mu}\thinspace,
\end{equation}
and $g^{\prime}$ can also be related to the electric charge $g^{\prime}=e/\cos\theta$.
Since we are ultimately interested in the consequences of the interaction
in Eq.\,\eqref{eq:Seff4} in the very low energy regime, contributions
involving the $Z_{\mu}$ field (which is very heavy) can be disregarded.
The relevant effective interaction thus generated is
\begin{equation}
S_{\mathrm{photon}}\left[A\right]\supset C\, e\,\phi\epsilon^{\rho\mu\nu\lambda}b_{\lambda}d^{\kappa}F_{\rho\mu}F_{\kappa\nu}\,.\label{eq:Seff4a}
\end{equation}
To recognize in \eqref{eq:Seff4a} an expression similar to \eqref{eq:axion-photon-coupling},
one has to note that 
\begin{equation}
\epsilon^{\rho\mu\nu\lambda}b_{\lambda}d^{\kappa}F_{\rho\mu}F_{\kappa\nu}=2\left(b\cdot d\right)\vec{E}\cdot\vec{B}\,,\label{eq:simplification1}
\end{equation}
therefore 
\begin{equation}
S_{\mathrm{photon}}\left[A\right]\supset2C\, e\left(b\cdot d\right)\,\phi\vec{E}\cdot\vec{B}\,.\label{eq:Seff5}
\end{equation}
This is the main result of this paper. We obtain with the mechanism
described in the previous paragraphs a term of the same form as Eq.\,\eqref{eq:axion-photon-coupling},
which is the most relevant interaction involving the pseudo-scalar,
from the point of view of current experimental searches. 

We can now argue that the interaction in Eq.\,\eqref{eq:Seff5} is
indeed the dominant one in the low energy regime. It will be essential
the assumption that the fermion mass $m$ in Eq.\,\eqref{eq:Lagrangian}
is very high: notice however that the resulting low energy interaction\,\eqref{eq:Seff5}
is independent of the mass scale $m$, differently from what happens
in LI theories as discussed after Eq.\,\ref{eq:leff}, in which the
coupling $g_{\phi\gamma}\sim1/f_{a}$ ($f_{a}$ being the relevant
large mass scale in that case). The mass independence of Eq.\,\eqref{eq:Seff5}
means that we can consider $m$ as large as necessary for the following
arguments to hold, without modifying the intensity of the generated
ALP-photon interaction.

Consider for example the second and third terms in the right hand
side of Eq.\,\eqref{eq:ourPi2}. These could induce higher derivatives
operators with nontrivial dependence on the directions of the LV parameters
$b^{\mu},\, d^{\mu}$. The coefficient $E$, whose explicit form is
\begin{align}
E= & \frac{i}{2\pi^{2}}\int_{0}^{1}dz\int_{0}^{1-z}dx\,\frac{4x^{3}-5x^{2}+x}{x\left(x-1\right)k^{2}+m^{2}}\nonumber \\
 & -\frac{i}{2\pi^{2}}\int_{0}^{1}dz\int_{0}^{1-z}dx\,\frac{k^{2}x^{3}\left(x-1\right)^{2}-m^{2}x\left(x^{2}-1\right)}{\left[x\left(x-1\right)k^{2}+m^{2}\right]^{2}}\,,
\end{align}
would correspond to an interaction proportional to $\partial^{\mu}\phi\left(b_{\sigma}\partial_{\alpha}F^{\nu\sigma}\right)\left(d^{\rho}F_{\nu\rho}\right)$
in the low energy effective action. The integrals over the Feynman
parameters can be explicit performed and, assuming $m^{2}\gg k^{2}$,
we verify that 
\begin{equation}
E\sim-\frac{i}{12\pi^{2}m^{2}}\,,
\end{equation}
meaning the corresponding higher derivative term is strongly suppressed
in the low energy regime. In the same limit, it can be shown that
\begin{equation}
D=\frac{i}{2\pi^{2}}\int_{0}^{1}dz\int_{0}^{1-z}dx\,\frac{x\left(x-1\right)}{x\left(x-1\right)k^{2}+m^{2}}\sim-\frac{i}{24\pi^{2}m^{2}}\,.
\end{equation}
The same should happen to the remaining terms in Eq.\,\eqref{eq:ourPi}:
they should represent higher derivative terms, strongly suppressed
by powers of the inverse mass of the fermion, which is assumedly very
high.

\section{\label{sec:Implication-for-ALPs}Implication for ALPs Phenomenology}

ALPs are very light particles, and therefore the only relevant operators
in the phenomenology of these particles are the dominant ones in the
very low energy regime. We have argued, in the previous section, that
in this model the lowest dimensional, dominant operator is the one
given in Eq.\,\eqref{eq:Seff5}. It is surprising to notice that
even with the model starting with explicit Lorentz violation, this
interaction depends only on the scalar $b\cdot d=b^{\mu}d_{\mu}$%
\footnote{The classification of $b^{\mu}d_{\mu}$ as a \emph{scalar} might require
some clarification. It is important, when considering Lorentz violating
extensions of the Standard Model, to distinguish between observer
and particle Lorentz transformations. Under the first of these, all
Lorentz indices transform covariantly, while under the second, indices
of fields, derivatives and gamma matrices transform as usual, while
$b^{\mu}$ and $d^{\mu}$ remain constant. Therefore, $b^{\mu}d_{\mu}$
is a genuine Lorentz scalar under observer Lorentz transformation
because of the covariant contraction of indices, while it remain constant
under particle Lorentz transformation because of the constancy of
the $b^{\mu}$ and $d^{\mu}$ independently. Either way, the expression
$b^{\mu}d_{\mu}$ is invariant, which justifies our slight abuse of
nomenclature in simply calling it a \emph{scalar}.%
}. That means the underlying LV background involving the vectors $b_{\mu},d_{\mu}$
that we considered generated a \emph{Lorentz invariant} interaction,
which mimics precisely the effects of the ALPs interactions studied
in the Lorentz preserving context. We conclude that whenever the kind
of LV that we considered as the initial input for our calculations
turns out to happen in nature, the \emph{effective} ALP-photon coupling
that are measured by experiments is actually 
\begin{equation}
g_{\phi\gamma}^{\left(eff\right)}\phi F_{\mu\nu}\tilde{F}^{\mu\nu}=\left(g_{\phi\gamma}^{\left(LI\right)}+2C\, e\left(b\cdot d\right)\right)\phi F_{\mu\nu}\tilde{F}^{\mu\nu}\,,\label{eq:Geffetive}
\end{equation}
where $g_{\phi\gamma}^{\left(LI\right)}$ represents the contribution
generated by other -- Lorentz invariant -- interactions (such as the
ones due to the Peccei-Quinn mechanism for solving the strong CP problem,
described in Section \ref{sec:Introduction}). One might even say
that Lorentz violation could represent an alternative mechanism to
generate such couplings, that is, even if $g_{\phi\gamma}^{\left(LI\right)}=0$,
one could have the essential signals of the presence of an ALP-photon
interaction induced only by the LV.

Since the interaction in Eq.\,\eqref{eq:Seff5} is LI, one could
not resort to the usual signatures of LV theories for imposing constraints
on the LV parameters $b^{\mu},\, d^{\mu}$, such as a dependence on
the orientation of the laboratory to these fixed vectors. However,
if a QCD axion happens to be found, and the specific details of the
embedding of the PQ chiral symmetry in the Standard Model are fixed,
then small deviations from the mass and coupling relation derived
from Eq.\,\eqref{eq:mAgArelation} could be attributed to the effects
of the interaction in Eq.\,\eqref{eq:Seff5}, as can be seen in Eq.\,\eqref{eq:Geffetive}.
Conversely, bounds on the LV parameters $b^{\mu},\, d^{\mu}$ could
be drawn from a precise experimental verification of the QCD axion
predictions.

The appearance of the ambiguous constant $C$ in \eqref{eq:Seff5}
brings into question the technical puzzle involved in its calculation.
Assuming, for the sake of the argument, $C\sim1$, one roughly obtains
from \eqref{eq:Seff5} and \eqref{eq:axion-photon-coupling} the experimental
constraint 
\begin{equation}
e\left(b\cdot d\right)\lesssim10^{-10}\mbox{GeV}^{-1}\,.\label{eq:bounds}
\end{equation}
We are unaware of independent experimental constraints on the Lorentz
violating couplings $d^{\nu}F_{\mu\nu}\bar{\psi}\gamma^{\mu}\psi$
and $\phi\bar{\psi}\gamma_{5}\slashed{b}\psi$, but if one of these
could be found, we could put an experimental constraint on the other
one. Notice however that Eq.\,\eqref{eq:Seff5} is of second order
in LV parameters, so we might not be able to put very stringent bounds
on $b_{\mu}$ or $d_{\mu}$ based only on Eq.\,\eqref{eq:bounds}.
This is a question that deserves further study since it connects the
ambiguous constant $C$ with an experimentally observable quantity,
namely the effective coupling in Eq.\,\eqref{eq:Seff5}. We expect
that the ambiguity in the calculation of $C$ should be fixed by physical
constraints, maybe of the experimental nature.

\section{\label{sec:Conclusions-and-Perspectives}Conclusions and Perspectives}

In this work, we have shown that a model with LV interaction in the
high energy regime can induce, in the low energy limit, LI operators
that can be relevant to ALPs phenomenology. Our construction is based
on the assumption of the existence of a very massive fermion field,
with specific LV interactions, which upon integration will provide
suppression factors of inverse of its mass to all LV operators in
the photon effective action, except for the leading operator in Eq.\,\eqref{eq:Seff5},
which happens to be LI. Thus, we proposed a novel way to seek for
possible consequences of LV occurring in some fundamental high energy
theory, that is, looking at LI effects that could in principle be
seen in current experimental searches for light pseudoscalars. 

In summary, the connection between Lorentz violation and the phenomenology
of axion-like particles can have very interesting consequences both
for the searches of possible violations of standard spacetime symmetries,
and for the search of the elusive light pseudoscalars that could solve
many theoretical puzzles in our current understanding of the universe. 

To finish, one may wonder whether other relevant LI operators could
be induced by and underlying LV dynamics, maybe a LI operator relevant
to the Higgs phenomenology. We may consider the following model, 
\begin{equation}
{\cal L}=\bar{\psi}\left[i\slashed{\partial}-m-\gamma^{\mu}\left(g^{\prime}B_{\mu}+u^{\nu}\epsilon_{\mu\nu\lambda\rho}B^{\lambda\rho}\right)-\gamma_{5}\slashed{v}\varphi^{\prime}\right]\psi\,,\label{eq:higgs1}
\end{equation}
instead of Eq.\,\eqref{eq:Atilde}, where $u^{\nu}$ is a LV vector
and now $v_{\mu}$ is a LV pseudovector, and $\varphi^{\prime}$ a
real scalar field. Since $\psi$ is assumed to be a singlet under
the gauge symmetries of the Standard Model, it cannot directly couple
to the usual Higgs doublet field $H$. However, if there is a $H-\varphi^{\prime}$
interaction leading to a mixing after spontaneous symmetry breaking,
such that $\varphi^{\prime}=\langle\varphi^{\prime}\rangle+cos\alpha\,\varphi+sin\alpha\, h$,
with $\langle\varphi^{\prime}\rangle$ the $\varphi^{\prime}$ vacuum
expectation value and the unprimed fields being mass eigenstates,
then the Higgs field $h$ ends up coupled to $\psi$. An example of
interaction term leading to such a mixing is $H^{\dagger}H\,\varphi^{\prime2}$.
From Eq.\,\eqref{eq:higgs1}, by repeating the steps described in
the previous paragraphs, one would arrive at 
\begin{equation}
S_{\eff}\left[B\right]\supset2Cg^{\prime}\, F_{\rho\mu}\epsilon^{\rho\mu\nu\lambda}v_{\lambda}\epsilon_{\nu\kappa\sigma\tau}u^{\kappa}\,\varphi^{\prime}F^{\sigma\tau}\,.
\end{equation}
In this way one generates scalar couplings to the photon of the general
form 
\begin{equation}
\mathcal{L}_{h\gamma}=c_{1}\varphi^{\prime}\left(u^{\mu}F_{\mu\nu}\right)\left(v_{\alpha}F^{\alpha\nu}\right)+c_{2}\left(u\cdot v\right)\varphi^{\prime}F^{\mu\nu}F_{\mu\nu}\,.\label{eq:alternative-axionterm}
\end{equation}
The first term is sensitive to the direction of the Lorentz breaking
vectors $u_{\mu},v_{\mu}$, as can be made explicit by rewriting it
as 
\begin{equation}
c_{1}\varphi^{\prime}\left[-\left(u^{0}v^{0}\right)\vec{E}^{2}+\left(u^{0}\right)\vec{v}\cdot\left(\vec{E}\times\vec{B}\right)+\left(v^{0}\right)\vec{u}\cdot\left(\vec{E}\times\vec{B}\right)+\left(\vec{E}\cdot\vec{u}\right)\left(\vec{E}\cdot\vec{v}\right)+\vec{u}\cdot\left[\vec{B}\times\left(\vec{B}\times\vec{v}\right)\right]\right]\,.
\end{equation}
From this interaction, a typically Lorentz violating phenomenology
could be studied, namely, the dependence of physical measurements
on the direction of the vectors $v_{\mu},u_{\mu}$, which by assumption
are fixed in some preferred cosmological Lorentz frame.

The second term in Eq.\,\eqref{eq:alternative-axionterm}, however,
depend on the scalar $u\cdot v$, and with the mixing there appears
the interaction $c_{2}\left(u\cdot v\right)sin\alpha\, h\, F^{\mu\nu}F_{\mu\nu}$,
which is identical in form to the effective couplings in the Standard
Model responsible for the diphoton Higgs boson decay. This decay was
one of the main channels for the observation of the resonance at $126\,\mbox{GeV}$
recently discovered at the LHC\,\cite{AtlasHiggs:2012tfa,CMSHiggs:2012ufa}.
Being second order in the LV parameters, this correction should be
small enough for not to bring any obvious inconsistency with the observations,
however we might speculate whether LV could induce some small deviations
from the predictions of the Stardard Model that could be still measurable
in this or some other process. 

\bigskip{}

\textbf{Acknowledgements.} The authors acknowledges financial support
from the Brazilian agencies CNPq, under the processes 303438/2012-6
(A.Yu.P.), 471834/2011-4 (A.F.F.) and 302138/2010-2 (A.G.D.), and
FAPESP, under the process 2013/01231-6 (L.H.C.B.).

\end{document}